# Factors Modulating Software Design Quality

[1]Poornima U. S . [2] Suma. V.
*[1,2]Research and Industry Incubation Centre, DSCE*
*[1]RajaReddy Institute of Technology, Bangalore, India.*
*[2]Dayananda Sagar College of Engineering*
*Bangalore, India.*
[1] uspaims@gmail.com, [2]sumadsce@gmail.com

*Abstract*— Object oriented approach is one of the popular software development approach for managing complex systems with massive set of requirements. Unlike procedural approach, this approach captures the requirements as set of data rather than services. Further, class is considered as a key unit of the solution-domain with data and services wrapped together, representing architectural design of a basic module. Thus, system complexity is directly related to the number of modules and the degree of interaction between them. This could be mapped as a functional diagram with cardinalities between the modules. However, complexity is always a threat to quality at each stage of software development. Design phase is therefore one of the core influencing phases during development that selects the right architecture based on the problem statement which is bound to be measured for quality. Hence, software industries adapts several organization-specific principles, domain-specific patterns, metric standards and best practices to improve and measure the quality of both process and product. The paper highlights the factors which influence the overall design quality and metric's implication in improving the quality of final product. It also presents the solution domain as an interdependent layered architecture which has a greater impact on concluding the quality of the end product. This approach of design is a unique contribution to the domain of Object Oriented approach of software development. It also focuses on design metrics which ensures the implementation of right choice of design towards the retention of quality of the product.

## I. INTRODUCTION

Software industry is a business icon in the era of automation. It created an open platform inviting other industries to merge their business to serve the customer in a better manner. The problem domain addressed so far includes simple business to E-commerce, stand-alone to space technology, robot to embedded and numerous. The industry is eager to support the business and science by developing domain-specific quality software. The complexity of the problem statements varies with different domains, however the final goal of all software companies is to reduce the defect rate and increase the total quality [1][2]. Quality, a prime factor in software development, is a measurable unit need to be applied to both process strategy and end-product to sustain total customers satisfaction in the competitive market.

*Tolal_Quality=∑Q(Process)+Q(Product)           Eq.1.*

To attain total quality, industries adapt their own process strategy and measures which are project-specific and defined using past experience. Programming–in-the-large is the most commonly used development strategy in recent years for problem domains with variety of data-centric requirements. Such domains usually demands different development approach rather than traditional procedural methodology. The development process is not only focusing on the work product but also its reusability and extensibility for future reference and enhancement. It well supports the concept of modularity, making the system more flexible and maintainable for changes in future. The user defined data structure, a Class, imbibes the quality of modularization providing interrelated data and functions together. However, classes in isolation will not provide the higher level services. This can be achieved by defining the relationship between classes and objects in a solution model.

Industries imbibe several design principles for modeling the solution space of a data-oriented system. They serve as a key note for an architect for designing a basic unit class to package, contributing to overall design quality. Several such principles are proposed in the literature and combination of the facts present in these design principles result in different design patterns. The architectural model includes the patterns suitable for the problem statement. The quality of the end product is also influenced by the patterns and frameworks used in the solution domain since the complexity of the design depends on the degree of linkage between the elements present in the solution domain.

Plenty of measures are proposed to measure the quality of overall development process from requirement gathering to product testing.

The metrics defined proposes a hypothetical measure of different phases of software development. Overall process quality is cumulative quality of each process phase. [3][4][5].

$$Q(Process) = \sum_{k=1}^{n}(Q(Phasek)) \qquad \text{Eq.2}$$

where *n* is number of phases in software development adapted by industry. Though several best practices and quality standards are introduced in industry to account the quality of process, the defect rate in the design phase is increasing with small scale to large scale systems.

## II. MODELLING OF OBJECT ORIENTED SOLUTION DOMAIN

In recent years, the problem statements are more complex and prone to changes in a near future. Domain-specific business applications to space applications are distributed and networked in nature, need to be highly modularized in the solution domain. In OO development, requirements are perceived in a realistic way and solution space is populated with set of classes and objects. In fact, the physical and logical collaboration between classes and objects defines the system architecture as a whole providing a baseline for client requirements. The system is physically modelled through class relationships like inheritance, aggregation and association, where as collaboration between objects exhibits the system behaviour [6][7].

### A. Inheritance, Aggregation:

Requirements in OO systems are perceived as collection of different data. The segregation of data is then done based on the commonality existing among them and classified as group of classes. Abstraction of each class is defined on data set in most generic way, to support reusability for further demands.

Inheritance is a concept of OO approach supporting reusability and extensibility. The classes under construction can use data/services of existing classes in the hierarchy. Reusability /extensibility can be either vertical (multilevel) or horizontal (multiple) in ladder of classes.

Inheritance upholds sharing and completes the solution space by interconnecting the modules. Aggregation exhibits the logical and physical containment of classes within to simplify the module architecture. The logical clustering of classes (objects) through aggregation is a design artefact for the developers to enrich the overall structure of the system.

### B. Association, Links

Association is a way of interrelating the classes which are functionally independent each other. Objects of different classes are floating within/among the modules through links established between them. It promotes client/server architecture, thereby sharing the services facilitating the developers to merge multifaceted requirements to provide higher level services to the customers.

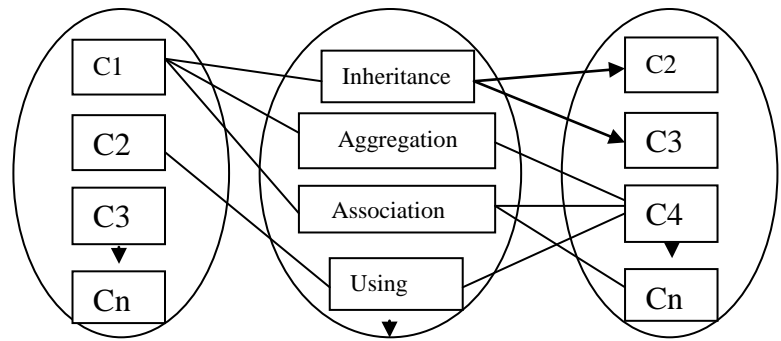

Figure 1. Relationship between classes in a solution domain.

### III. OBJECT ORIENTED DESIGN PRINCIPLES

A good design is always inherently flexible enough for modifications. Such designs are economic and less risky for future demands. Design is an art with no thumb rules, however, in OO development, few design principles for modeling classes, packages and interfaces are in practice [8][9][10][11][12[13].

*Class Design Principles:* Class is a basic building block of OO development approach. Success of the software design lies in well designed classes and their relationship in the solution space.

*a. Separation of Concern: Single Responsibility Principle:*

Modularization is the core concept of OO design. Module is designed with single responsibility/functionality to make the system more cohesive and easy adaptation of changes incurred. It also makes the system maintenance easier as changes in a module would not affect the rest of the system.

*b. Polymorphism: The Open Closed Principle:*

The operational attribute of a class, function/ method, defines the behavior of a subsystem. Floor of a class needed to be open for adding new behaviors without changing the existing source codes. It retains the logical independency between the modules, but extends the functionally as required by the client in future.

*c. Inheritance: The Liskov Substitution Principle:*

Extension and specialization of modules is an added advantage in OO approach. Logically not cohesive functionalities can be defined as a sub module or new module is defined which is substitutable to the parent class. This principle makes the architecture more realistic and scalable.

*d. Abstractions: The Dependency Inversion Principle:*

A good design is one which contains more abstract classes in the solution domain rather than concrete classes. Abstractions in the high level classes are more generic with scope to redefine it at the lower levels. However, this principle states that abstractions are not dependent on lower abstractions and encourage abstract classes than concrete one.

*e. The interface Segregation Principle:*

Interfaces are collection of functions with only signatures. Good design supports many client specific interfaces than generic one. Thus, a loaded class with functions is segregated as number of interfaces specific to the client requirements.

*Package Design Principles:* Applications are simply a network of group of classes, namely packages. However choosing a right class for a package (cohesion) and the dependency between packages (coupling) is a major contributing factor for design quality.

*Package cohesion principle:* Grouping of classes done based on the common platform, simplifies the architecture of a system. The classes in a group/package are cohesive each other to increase the design quality. To select a class for a right package, the package cohesion principles help the architect.

*f. The Release Reuse Equivalency Principle (RREP):* The core concept of OO technique is reusability of well defined classes/packages. Architects choose the classes with commonality in to a package. The release management trace-out the version of the class and put in the repository for reuse by the developer in future.

*g. The Common Closure Principle (CCP):*

Change is inevitable in software development. Classes having tendency to change together are grouped together to ease the future enhancement of the system.

*h. The Common reuse Principle (CRP):*

This principle states that the classes that are not reused together should not be in the same package. It supports 'separation of concern' principle with the package for better scalability and maintainability.

*Package coupling principle:* Dependency between the packages is essential in order to support reusability and effective project management. The cross-coupling between the existing packages should not hamper the overall system quality. The packages coupling principles is an insight for an architect before networking them.

*i. The Acyclic Dependencies Principle (ADP):*

This principle states that the dependencies between the packages should not form any cycles. Since packages are granules for release, its functionality and dependency with other packages are thoroughly analyzed to increase the stability of system design.

Transitive dependency with package cycles results in bad design.

*j. The Stable Dependencies Principle(SDP):*

A stable package is one which has all incoming dependencies, but no outgoing dependencies. Such packages are independent and changes in other part of the system will not affect them. However, an instable package contains all outgoing dependencies and is very much dependent on the packages to which it is connected. Instability factor of a package can be measured by value of its afferent and efferent coupling ratio.

Instability $I = \frac{Ce}{ca+ce}$  $0 >= I <= 1$  Eq.3.

Ce (Efferent coupling) provides a count on outgoing dependencies,
Ca (Afferent coupling) provides a count on incoming dependencies.

The package will be instable when there is incoming dependencies.

*k. The Stable Abstractness Principle (SAP):*

The package could be listed as stable when other packages are dependent on it. But such packages become too rigid as it would not support the changes, if so, would have ripple effect on the dependent packages. However by making such packages as abstract, the designer can extend the package rather than doing changes in it. Thus SAP supports both stability and extensibility factors of a good design.

The abstractness of a package is depending on the number of abstract classes it posses. The metric for abstractness

$A = \frac{Na}{Nc}$ where  Eq.4.

$Na$ stands for number of abstract classes and $Nc$ is total number of classes in a package.

IV. DESIGN PATTERNS, FRAMEWORKS AND SOFTWARE QUALITY FACTORS

Software design patterns are templates for commonly occurring problem in a software design. It describes the form of Classes, Objects, Inheritance, Aggregation and the communication between the objects and classes for a particular context. The framework is a collection of such patterns for a problem statement defining the high level abstraction and good communication between designer and user. Several design patterns are in practice and a good pattern is one which makes the overall software design more flexible, elegant and reusable. The key factors play important role in quality design in very pattern is coupling and cohesion. They express the strength within and among the classes present in a solution scenario [14][15][16].

*A. Cohesion*

Cohesion represents the degree of bondage between the elements in a class. High cohesive class posses the elements much related to each other and otherwise the class would be decomposed in to subclasses with interrelated members. The grouping of members is done based on the similar property they posses or they do on set of common data. A well designed class is highly cohesive if all the members focus on the same functionality. Several metrics are in practice to measure the cohesiveness of a class in a solution domain.

*1. Lack of Cohesion in Methods (LCOM):*

It measures the quality of a class in a solution domain. Cohesion refers the degree of interconnectivity between attributes of a class. A class is cohesive if it cannot be further divided in to subclasses. It measures the method behaviour and its relevance where it is defined. Pair of methods using data object proves the cohesiveness where as the methods not participating in data access makes it less cohesive. Consider C is a class and M1,M2...Mn are its methods using set of class instances. I1={a,b,c,d}, I2={a,b,c} and I3={x,y,z} are set of instances used by the methods M1,M2 and M3 respectively. If intersection of object set is non-empty then the method using them is cohesive and their relevance in the class is proved. i.e. I1 ∩ I2 = {a, b, c} means M1 and M2 are cohesive. But intersection of I1, I3 and I2, I3 is empty set. High count in LCOM shows less cohesiveness and class need to be divided to subclasses.

### 2. Weighted Methods per Class (WMC)

It measures the complexity of a class- operational attribute, methods, in terms of effort and time for development and maintenance. Complexity of a class is a cumulative sum of complexity of all its methods. The objective is to keep it low to uphold design quality

$$WMC(C) = \sum_{I=0}^{n} \binom{n}{k} Ci(Mi) \qquad \text{Eq.5.}$$

### B. Coupling

Sub modules in isolation would not provide the high level services. However, the coupling between the methods serves the user with rich set of operations. Coupling is the interdependency of sub modules within a system either at the class or object level. It supports reusability, extensibility also eases scalability and maintainability [17][18][19].

#### 1. *Class level coupling:*

The design domain of a problem statement contains numerous classes as representative of requirements set. The static connectivity between the classes either horizontality or vertically influences the design quality. Few metrics are proposed in the literature for coupling by Chidamber and Kemener, which measures the dependency between the modules.

#### 1a) Depth of Inheritance Tree (DIT)

It measures the *vertical* growth of a class. Inheritance supports reusability; however the complexity is directly proportional to the distance between leaf and parent class. Deeper tree structure is prone to higher complexity as it is difficult to access end class behaviour.

#### 1b) Number Of Children (NOC)

It is a metric to measure the *horizontal* growth of a class. The Immediate subclasses in a hierarchy show the greater reusability. System functional quality is highly dependable on abstractness of the parent class. Much effort is required in testing if tree grows in both directions.

#### 2. *Object level coupling*

#### 2a) Coupling Between Object classes (CBO)

It measures the interdependency between the classes. An object of a class can use the service or object of another class. The objective is to reduce the much interdependency (cross coupling) to increase the clarity of the solution. However, the effective coupling between the classes are depending on the language constructs and the pattern framework chosen by the architect [20][21][22][23].

## V. LAYERED ARCHITECTURE

Design of a solution space is a collection of set of different activities carried out in a sequential manner. Each set of activities can be brought under a layer, delivering the respective work products. The quality/flaws of the deliverables propagate from one layer to another, affecting the total system quality.

Hence, in object oriented methodology, each class need to be designed at most careful as a basic step of design since it builds the system strong and flexible for future use. The key principles for defining the class and establishing the relationship among them contribute the design quality considered as major steps (layers) for system design. These relationships, coupling and cohesion are the prime attributes for measuring system quality. To measure it, metrics are in practice as a quality quantifiers which provides the design quality at different levels of design.

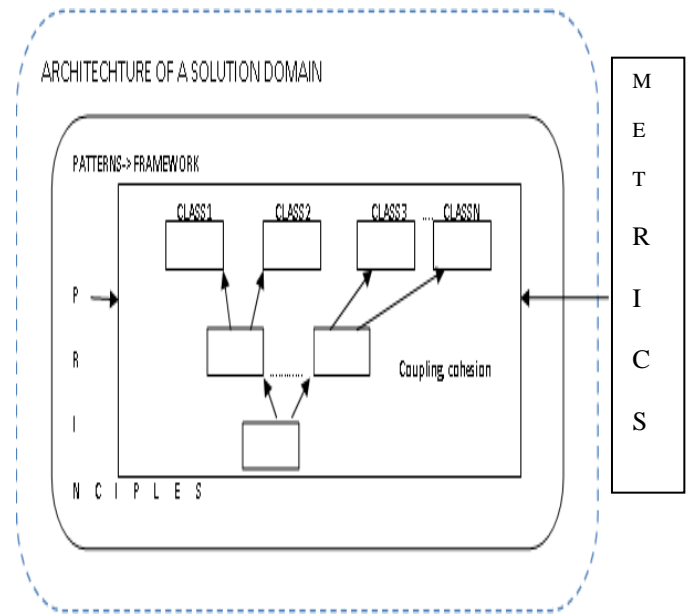

Fig. 2. The layered architecture for Object Oriented solution domain.

## VI. CONCLUSION

Object Oriented methodology is a recent trend in software development for addressing huge problem domain with verity of requirements. The success of such projects relies on the quality of overall design from modelling till quality assessment metrics. Though various patterns are available, it is the architect's cognitive ability to mix and match the patterns to suit the current problem statement in hand. The overall design quality basically depends on the complexity of the dependency between the modules. Since design quality cannot be achieved in a single step, quality of each step in a design activity architecture, as depicted in Fig.2. contributes to overall design quality. Coupling and cohesion as a result of relationship between classes are significant factors that influences design quality. Metrics proposed for measuring coupling and cohesion would be strengthening more to provide an empirical estimation for the designer on design quality. As flawless software is the destiny for developer and customer, our future work would provide better magnitude for software quality to retain total customer satisfaction in the market.

The uniqueness of this paper is further to introduce a layered approach of object oriented mode of software development. This paper limits towards its introduction and our forth coming work proves with data and design support for the same.